\documentclass[aps,pre,twocolumn,groupedaddress,showpacs]{revtex4}
\usepackage{epsf}
\newcommand{\lab}[1]{\label{#1}}

\newcommand{\ci}[1]{\cite{#1}}
\newcommand{\be}{\begin{equation}}
\newcommand{\ee}{\end{equation}}
\newcommand{\ba}{\begin{eqnarray}}
\newcommand{\ea}{\end{eqnarray}}
\date{}

\begin{document}
    \title{QUANTUM CHAOS IN THE YANG-MILLS-HIGGS SYSTEM AT FINITE TEMPERATURE}

\author{D.U.Matrasulov$^{(a)}$,
F.C.Khanna$^{(a)}$, U.R.Salomov$^{(b)}$ and  A.E.Santana$^{(c)}$
\thanks{e-mail: lps@infonet.uz}}
\date{\today}
\affiliation{
(a) Physics Department University of Alberta\\
 Edmonton Alberta, T6G 2J1 Canada\\
 and TRIUMF, 4004 Wersbrook Mall,\\ Vancouver, British Columbia, Canada, V6T2A3\\
(b) Heat Physics Department of the Uzbek Academy of
Sciences,\\
28 Katartal St.,700135 Tashkent, Uzbekistan
(c) Instituto de Fisica,
  Universidade be Brasilia, 70910-900, Brasilia, DF, Brazil}

\begin{abstract}
The quantum chaos in the finite-temperature Yang-Mills-Higgs system  is studied.
The energy spectrum  of a spatially homogeneous $SU(2)$ Yang-Mills-Higgs
is calculated within thermofield dynamics.
 Level statistics
 of the  spectra is studied by plotting nearest-level spacing distribution histograms.
 It is found that finite temperature effects lead to a strengthening of chaotic effects, i.e. spectrum which has Poissonian distribution at
 zero temperature has Gaussian distribution at finite-temperature.
\end{abstract}
\maketitle
\newpage
\section{Introduction}

Quantum chaos is a relatively new area in physics and has been the subject of extensive studies
for the  last two decades \ci{Brod}-\ci{Zas}.
It has found applications in atomic and molecular physics, nuclear physics and  condensed matter physics.
In the past few years there is growing interest in quantum chaos in particle physics, too.
Being the quantum theory of classically chaotic systems the quantum chaology studies
fluctuations in the energy spectra and
wave functions of such systems.
It is well known, that  the energy spectra of systems, whose classical counterparts are chaotic,
has the same statistical properties
as those for random matrices. Therefore one of the main topics in the quantum chaology is to study
the statistical properties
of the classically chaotic systems.
Recently energy fluctuations and quantum chaos in hadrons and QCD has become a subject of theoretical studies \ci{Gu}-\ci{Buk}.
In particular, it is  found that the quark-gluon system in QCD is governed by quantum chaos in both confined and deconfined
phases \ci{Hal,Bit}.
The statistical analysis of the measured meson and baryon spectra shows that there is quantum chaos phenomenon
in these systems \ci{Pasc}.
The  study of the charmonium spectral statistics  and its dependence on color screening has established quantum chaotic behaviour
\ci{Gu}. It was claimed that such a behaviour could be the reason for  $J/\Psi$ suppression \ci{Gu}.

In recent years there has been  considerable interest to determine the role of dynamical  chaos in field theories \cite{sav83}-\cite{mat97}.
Chaotic properties of Yang-Mills\cite{sav83,sav85}, Yang-Mills-Higgs
\cite{sav85}-\cite{sal95a} and Abelian Higgs \cite{kaw95}
have been treated. The main point in this considerations is the fact that the Hamiltonians
of the Yang-Mills and Yang-Mills-Higgs systems can be written in the same form as those for the coupled nonlinear oscillators.
This allows one to use, for their treatment the same methods as in the case of coupled nonlinear oscillators.
Quantum chaos in Yang-Mills-Higgs system was also studied recently \cite{sal97,sal99}. However all  works on chaos in  field theories and hadrons  are restricted to considering zero-temperature cases.

In this paper we study quantum chaos at finite-temperature in the Yang-Mills-Higgs system.
Recent advances in heavy ion collision experiments  allow one to create hot and dense hadronic and quark-gluon matters.
The role of finite-temperature effects in such  systems becomes important.
Especially  in quark-gluon or nuclear matter hadrons behave as complex systems
where  strong level fluctuations can be observed.
Thus  the role of finite-temperature effects as well as level fluctuations  are crucial in such a systems,
that obviously leads to a need for studying quantum chaos at finite-temperature.
In particular, finite-temperature effects cause fluctuations in their  energy spectra.

We address the problem of heat-bath in quantum chaos through the thermofield dynamics (TFD) formalism, a real time
finite temperature field theory \cite{um,tak,das,gui,rakh}. TFD is interesting for our proposal here by its remarkable algebraic
structure (this is not the case of the Matsubara \cite{mats} or the Schwinger-Keldysh \cite{keld} formalisms). Actually the central ideas
of TFD involve an algebraic doubling in the degrees of freedom and a Bogoliubov transformation giving rise to the
thermal variables. As we will show, this TFD prescription is a useful tool to explore finite-temperature effects in the
energy fluctuations of the Yang-Mills-Higgs Hamiltonian, represented in terms of the annihilation and
creation operators.

In section II we present the Yang-Mills-Higgs system at zero temperature. In section III the Yang-Mills-Higgs system at finite-temperature
is studied using the TFD formalism.
In section IV we present numerical results for finite-temperature and compare them to zero-temperature results, thus bringing out
crucial role of finite-temperature in Yang-Mills-Higgs system and  quantum chaos. Finally in Section V we present some conclusions
and new directions to pursue the important question of quantum chaos.

\section{Zero temperature case}
The Lagrangian for Yang-Mills-Higgs system  with $SU(2)$ symmetry  is given as
\be
L= -\frac{1}{4}F^a_{\mu\nu}F^{\mu\nu}_a +\frac{1}{2}(D_{\mu}\phi)^+ (D^{\mu}\phi) -V(\phi)
\ee
where
$$
F^a_{\mu\nu} = \partial_{\mu}A^{a}_{\nu} - \partial_{\nu}A^{a}_{\mu}  +g A^{b}_{\mu}A^{c}_{\nu}
$$
$$
(D_{\mu}\phi) = \partial_{\mu}\phi -igA^{b}_{\mu}T^b \phi
$$
with $T^b =\sigma/2$, $b=1,2,3$ generators of the $SU(2)$ algebra and $g$ is a coupling constant.
The potential of the scalar (Higgs) field is
$$
V(\phi) =\mu^2|\phi|^2 +\lambda |\phi|^4,
$$
where $\mu$ and $\lambda$ are constants.
Here we give a brief  description of the non-thermal case  \cite{sal97}.
In $(2+1)$-dimensional Minkowski space and for spatially homogeneous Yang-Mills and Higgs fields which satisfy the conditions
$$
\partial_i A^a_{\mu} =\partial_i\phi = 0, \;\;\;\;\; i =1,2,;
$$
and in the gauge $A_0^{\alpha} =0$, the Lagrangian can be written as
$$
L = \dot{\vec\phi^2} +\frac{1}{2}(\dot{\vec A_1^{2}} +\dot{\vec A_1^{2}} )- g[\frac{1}{2}\vec A_1^{2}\vec A_2^{2}
 -\frac{1}{2}(\vec A_1\vec A_2)^2+
$$
\be
(\vec {A_1^2}+\vec A_2^2)\vec\phi^2 - (\vec A_1\vec\phi )^2 -  (\vec A_2\vec\phi )^2] - V(\vec\phi),
\lab{lag2}
\ee
where $\vec\phi = (\phi^1, \phi^2, \phi^3)$, $\vec A_1 = (A_1^1, A_1^2, A_1^3)$ and  $\vec A_2 = (A_2^1, A_2^2, A_2^3)$

To treat the chaotic dynamics it is convenient to use the canonical formalism and work with the Hamiltonian of the system instead of
the Lagrangian. The Hamiltonian of the system  can be written as\cite{sal97}
\be
H =\frac{1}{2}(p_1^2 +p_2^2)+g^2v^2(q_1^2 +q_2^2) +\frac{1}{2}g^2q_1^2 q_2^2,
\ee
where
$\vec \phi_0 =(0,0,v)$ $q_1=A_1^1, \; q_2=A_2^2$ (other components of the Yang-Mills fields are zero) $p_1 =\dot q_1$ and $p_2 =\dot q_2$, with
$\omega^2 =2g^2v^2$  being the mass term of the Yang-Mills field.
This is the Hamiltonian of the classical system.
Replacing  $p_i$ and $q_i$  by operators  and introducing  the following creation and destruction operators
$$
\hat a_k =\sqrt{\frac{\omega}{2}}\hat q_k +i\sqrt{\frac{1}{2\omega}}\hat p_k\;\;\; \hat a^+_k =\sqrt{\frac{\omega}{2}}\hat q_k -i\sqrt{\frac{1}{2\omega}}\hat p_k
$$
we obtain
the corresponding quantum Hamiltonian:
\be
H   = H_0 +\frac{1}{2}g^2V
\label{ham}
\ee
 where
 $$
 H_0 =\omega (a_1 a^+_1 +a_2 a^+_2+1),
 $$
and
$$
V = \frac{1}{4\omega^2} (a_1 +a_1^+)^2(a_2 +a_2^+)^2 ,
$$

with $\omega^2 =2g^2v^2$ and the operators $\hat a_k$ and $\hat a_l^+$ satisfy the commutation relations
$[\hat a_k, \hat a_l^+] =\delta_{kl},\;\; k,l=1,2.$
The  eigenvalues of this Hamitonian are calculated by numerical
diagonalization  of the truncated matrix of the quantum  Yang-Mills-Higgs Hamiltonian in the basis
of the harmonic oscillator wave functions \cite{sal97}.
The matrix elements of $H_0$ and $V$ are
 $$
<n_1',n_2'|H_0| n_1,n_2> = \omega(n_1+n_2+1)\delta_{n_1'n_1}\delta_{n_2'n_2},
 $$
and
$$
<n_1',n_2'|V| n_1,n_2> =\frac{1}{4\omega^2} \{\sqrt{n_1(n_1-1)}\delta_{n_1' n_1-2}+
$$
$$
\sqrt{n_1(n_1-1)}\delta_{n_1' n_1+2}+ (2n_1+1)\delta_{n_1' n_1}\}
$$
$$
\times \{ \sqrt{n_2(n_2-1)}\delta_{n_2' n_2-2}+
$$
$$
\sqrt{n_2(n_2-1)}\delta_{n_2' n_2+2}+ (2n_2+1)\delta_{n_2' n_2} \}
$$
The numerical procedure for diagonalization of this matrix is described by Salasnich \ci{sal97}.
We use the same method in the case of finite-temperature calculations.
\section{Finite-temperature case}

To treat quantum chaos in the  finite-temperature Yang-Mills-Higgs system, we apply Thermofield Dynamics(TFD).
TFD is a real time operator formalism of quantum field theory at finite temperature in which any physical
system can be constructed  from a temperature dependent vacuum $|0(\beta)>$ which is a pure state.
The thermal average of  any operator is equal to the expectation value between the pure vacuum state $|0(\beta)>$
defined by applying Bogolyubov transformations to the usual vacuum state .
Furthermore,
TFD has two main features. The first one is the  doubling of the Fock space such that  the original
Fock space and its double are defined non-tilde and tilde space respectively. All operators are also doubled
and the finite-temperature
creation and annihilation operators are constructed by  Bogoluybov transformation between tilde and non-tilde operators.
This is the same procedure in writing down the vacuum state at finite temperature.
Mathematically, the field operators have the following properties:
$$
{(A_i  A_j\tilde )} = \tilde{A_i}\tilde{A_j},
$$
$$
(cA_i + A_j\tilde) =c^*\tilde{A_i} + \tilde{A_j},
$$
$$
(A_i^*\tilde) =(\tilde{A_i})^+,
$$
$$
(\tilde{A_i}\tilde) =A_i
$$
$$
[\tilde{A_i}  A_j] =0.
$$

The Yang-Mills-Higgs Hamiltonian in TFD is given as
\be
\hat H =H- \tilde H
\ee
where $H$ is given by eq.\ref{ham} and $\tilde H$ is given as
$$
\tilde H   = \tilde H_0 +\frac{1}{2}g^2\tilde V
$$
with

 $$
\tilde  H_0 =\omega (\tilde a_1 \tilde a^+_1 +\tilde a_2\tilde a^+_2+1),
 $$

and
\be
\tilde V =\frac{1}{4\omega^2} (\tilde a_1 +\tilde a_1^+)^2(\tilde a_2 +\tilde a_2^+)^2,
\ee

 It has been established \cite{Ademir} that in an algebraic approach the
doubled set of operators may be considered as a set of operators that relate to the physical observables, $O$,
and a second set that are generators of symmetries, $\hat O$.  
The hat operators are responsible, in particular, for time development
and are needed for the purpose of scattering, decay and any transitions between states.
The physical observables lead to the quantities that are measured in experiment.
Both for physical observables and generators of symmetry after Bogolyubov transformations leading to finite-temperature
creation and annihilation operators and to a pure vacuum state only the non-tilde operators are required for getting
the appropriate matrix elements. However, it is clear that for an analysis of any system at finite temperature
both set of operators , $O$ and $\hat O$, are needed since it is important to generate the appropriate
symmetry of the system in any time development while considering the
matrix element we have decided to investigate the chaotic  behaviour for the Hamiltonian
that relates to physical observables, $H$ and the Hamiltonian that generates the symmetry of the system, $\hat H$.
Results of the finite-temperature quantum chaos in Yang-Mills-Higgs system will be displayed for the physical observables, and for the
generators of symmetry, $\hat H$.
 Only then we  will draw conclusions
about the approach of quantum chaotic behaviour in the finite-temperature quantum field theory case of Yang-Mills-Higgs theory.

First we need to rewrite the non-tilde part of the Hamiltonian
 in the temperature-dependent form using the Bogolyubov transformations which are given by
$$
a_k =a_k(\beta)cosh\theta +\tilde a^+_k(\beta)sinh\theta
$$
$$
a^+_k =a^+_k(\beta)cosh\theta +\tilde a_k(\beta)sinh\theta
$$
$$
\tilde a_k =a^+_k(\beta)sinh\theta +\tilde a_k(\beta)cosh\theta
$$
$$
\tilde a^+_k =a_k(\beta)sinh\theta +\tilde a^+_k(\beta)cosh\theta
$$

where
$$
\beta = \frac{\omega}{k_BT}
$$
where tilde and non-tilde creation and annihilation operators satisfy the following commutation relations:
$$
[a_k(\beta), a_l^+(\beta)] =\delta_{kl}\;\;\;\; [\tilde a_k(\beta), \tilde a_l^+(\beta)] =\delta_{kl}
$$
$l,k = 1,2$, and
$sinh^{2}\theta =(e^{\beta}-1)^{-1}.$
All other commutation relations are zero.

Then the  temperature-dependent forms of $H_0$ and $\tilde H_0$ are
$$
H_0 = \omega \{ (F_1+F_2)cosh^2 \theta +
$$
$$
(L_1+L_2)sinh^2\theta + (S_1+S_2)cosh\theta sinh\theta +1  \},
$$
$$
\tilde H_0 = \omega \{ (F_1+F_2)sinh^2 \theta +
$$
$$
(L_1+L_2)cosh^2\theta + (S_1+S_2)cosh\theta sinh\theta +1  \},
$$

where
$$
F_k = a^+_k (\beta)a_k(\beta),
$$
$$
L_k =  \tilde a_k(\beta)\tilde a^+_k(\beta),
$$
$$
S_k = a^+_k(\beta)\tilde a^+_k(\beta) +\tilde a^+_k(\beta)a_k(\beta),
$$

For $V$ and $\tilde V$ we have
$$
V=\frac{1}{4\omega^2}\{A_1cosh^2\theta +B_1cosh\theta sinh\theta +
$$
$$
C_1sinh^2\theta\}\{A_2cosh^2\theta +
B_2cosh\theta sinh\theta +C_2sinh^2\theta\}
$$
$$
\tilde V=\frac{1}{4\omega^2}\{A_1sinh^2\theta +B_1cosh\theta sinh\theta +C_1cosh^2\theta\}\times
$$
$$
\times\{A_2sinh^2\theta +
B_2cosh\theta sinh\theta +C_2cosh^2\theta\},
$$
where
$$
A_k = (a_k(\beta)+ a^+_k(\beta))^2,
$$

$$
B_k =(a_k(\beta)+ a^+_k(\beta)) (\tilde a^+_k(\beta)+ \tilde a_k(\beta)) +
$$
$$
(\tilde a^+_k(\beta)+ \tilde a_k(\beta))(a_k(\beta)+ a^+_k(\beta)),
$$
and
$$
C_k =  (\tilde a^+_k(\beta)+ \tilde a_k(\beta))^2.
$$

In the first approach the energy eigenvalues of the thermal Yang-Mills-Higgs system can be calculated by diagonolization of the
following  matrix:
\be
R=<n_1'n_2',\tilde n_1'\tilde n_2'| H_0+\frac{1}{2}g^2 V |n_1n_2,\tilde n_1\tilde n_2>.
\label{ar}
\ee
The elements of the matrix can be calculated explicitly:

$$
<n_1'n_2',\tilde n_1'\tilde n_2'| H_0 |n_1n_2,\tilde n_1\tilde n_2>=
$$
$$
=\omega \{(n_1+n_2+1)(1+2sinh^2\theta) \delta_{n_1'n_1}\delta_{n_2'n_2}+
$$
$$
\{n_1\delta_{n_1'n_1-1}\delta_{n_2'n_2} +n_2 \delta_{n_2'n_2-1}\delta_{n_1'n_1}  +(n_1+1)\delta_{n_1'n_1+1}\delta_{n_2'n_2} +
$$
$$
(n_2+1)\delta_{n_2'n_2+1}\delta_{n_1'n_1} )\}cosh\theta sinh\theta\}
$$
and for $V$
$$
<n_1'n_2',\tilde n_1'\tilde n_2'| V |n_1n_2,\tilde n_1\tilde n_2>=
$$
$$
\frac{1}{4\omega^2}<n_1'n_2',\tilde n_1'\tilde n_2'|\{A_1A_2cosh^4\theta +C_1C_2sinh^4\theta  +
$$
$$
(A_1C_2 +B_1B_2 +C_1A_2)cosh^2\theta sinh^2\theta +
$$
$$
(A_1B_2 +B_1A_2)cosh^3\theta sinh\theta +
$$
\be
(B_1C_2 +C_1B_2)cosh\theta sinh^3\theta\}|n_1n_2,\tilde n_1, \tilde n_2>
\label{matx1}
\ee

The matrix elements of $A_k, B_k$ and $C_k$ are given as
$$
<n_1'n_2',\tilde n_1'\tilde n_2'| A_k |n_1n_2,\tilde n_1\tilde n_2> =$$
$$
= \sqrt{n_k(n_k-1)}\delta_{n_k'n_k-2} +(2n_k+1) \delta_{n_k'n_k} +
$$
$$
+ \sqrt{(n_k+1)(n_k+2)}\delta_{n_k'n_k+2},
$$

$$
<n_1'n_2',\tilde n_1'\tilde n_2'| B_k |n_1n_2,\tilde n_1\tilde n_2>
$$
$$
=2n_k \delta_{n_k'n_k-1} + (2n_k+1) \delta_{n_k'n_k+1},
$$
and
$$
<n_1'n_2',\tilde n_1'\tilde n_2'| C_k |n_1n_2,\tilde n_1\tilde n_2> =$$
$$
= \sqrt{n_k(n_k-1)}\delta_{n_k'n_k-2} +(2n_k+1) \delta_{n_k'n_k} +
$$
$$
\sqrt{(n_k+1)(n_k+2)}\delta_{n_k'n_k+2}.
$$

The  calculation of the matrix element of $\hat H$, the generator of symmetry, gives us the following matrix:
$$
Z= <n_1'n_2',\tilde n_1'\tilde n_2'| H -\tilde H |n_1n_2,\tilde n_1\tilde n_2> =
$$
$$
= \omega \{-(2cosh^2\theta)\delta_{n_1'n_1}\delta_{n_2'n_2}+
$$
$$
+(n_1+n_2) \delta_{n_1'n_1}\delta_{n_2'n_2}\}+
$$
$$
+\frac{g^2}{2 \omega^2}cosh^4\theta \{\sqrt{n_1(n_1-1)}\delta_{n_1'n_1-2} +(2n_1+1) \delta_{n_1'n_1} +
$$
\be
+\sqrt{(n_1+1)(n_1+2)}\delta_{n_1'n_1+2}\}
\label{matx2}
\ee

Diagonalizing the matrices $R$ and $Z$  numerically we obtain the energy
eigenvalues of the Yang-Mills-Higgs system
for the Hamiltonians $H$ and $\hat H$ at finite-temperature.
As it was mentioned \ci{sal95a} the numerical energy levels depends on the
dimension of the truncated matrix. We compute the numerical levels in double
precision. The matrix dimension is $1156 \times 1156$, i.e. we calculate the first 1156 eigenvalues..
Then the statistical properties of the spectra are found. We use standard
unfolding procedure in order to remove
the secular variation of the level density as a function of the energy $E$,
for each value of the coupling constant the corresponding spectrum is mapped,
by a numerical procedure described in \ci{manfr}.

One of the main characteristics of the statistical properties of the spectra is
the level spacing distribution \ci{Eckh88,Gutz90} function.
In this work we calculate the nearest-neighbor level-spacing distribution
\ci{Eckh88,Gutz90,Guhr,Gu}.
The nearest neighbor level spacings are defined as $S_i = \tilde E_{i+1}- \tilde E_i$,
where $\tilde E_i$ are the energies of the unfolded levels, which are obtained by the following way:
The spectrum $\{E_i \}$  is separated into smoothed average part and fluctuating parts. Then the  number of the levels below $E$ is counted
and the following staircase function is defined:
$$
N(E)  = N_{av}(E) +N_{fluct}(E).
$$

The unfolded spectrum is finally obtained with the mapping
$$
\tilde E_i = N_{av}(E_i).
$$

Then the nearest level spacing distribution function $P(S)$ is defined as the probability of  $S$  lying within the  infinitesimal interval $[S, S+dS]$.

For the quantum systems which are chaotic in the classical limit this  distribution function is the same as that of the random matrices \ci{Eckh88,Guhr}.
For  systems which are regular in the classical limit its behaviour is close to a Poissonian distribution function.
This distribution is usually taken to be a Gaussian with a parameter $d$:
$$
P(H) \sim exp(-Tr\{HH^+\}/2d^2),
$$

and the random matrix ensemble corresponding to this distribution is called the Gaussian ensemble.
For Hamiltonians invariant under rotational and time-reversal transformations the  corresponding ensemble
of matrices is called the Gaussian orthogonal ensemble (GOE).
It was established \ci{Eckh88,Gutz90,Guhr} that GOE describes  the statistical fluctuation properties of a quantum system whose classical analog is
completely chaotic.
The nearest neighbor level spacing  distribution for GOE is described by  the Wigner distribution:
\be
 P(S) = \frac{1}{2}\pi S exp(-\frac{1}{4}\pi S^2).
 \lab{wig}
\ee
 The usual way to study  the level statistics is to compare the calculated nearest-neighbor level-spacing distribution histogram with the
 Wigner distribution.

 For systems whose classical motion is neither regular nor fully chaotic (mixed dynamics)  the level spacing distribution will be
 intermediate between the Poisson and GOE limits. Several  empirical functional forms for the distribution have been suggested for this case \ci{Guhr}.
If the Hamiltonian is not time-reversal invariant, irrespective of its behavior under rotations, the Hamiltonian matrices are  complex Hermitian and
the corresponding ensemble is called Gaussian unitary ensemble (GUE).
If the Hamiltonian of the system is time-reversal invariant but not invariant under rotations, then the corresponding ensemble is called the Gaussian
symplectic ensemble. In the next section we will study the numerical results for the level spacing and then analyze them to classify
them in one of the these categories. It is to be emphasized that our Yang-Mills-Higgs Hamiltonian is time-reversal and rotational invariant.
\section{Numerical results}
On diagonalizing the vacuum expectation value of the Hamiltonian $H$, physical observables, and $\hat H$, the generator of
symmetry, given in Eq. \ref{ar} and Eq. \ref{matx2}, respectively the spectra are analyzed by considering  the level-spacing distribution.
Then the criterion mentioned in the last section allows us to classify the system as chaotic or non-chaotic.
In Figs. 1 and 2 we plot the level spacing distributions for different values of parameters $\omega$ and $\theta$ for 
the energy spectrum calculated by diagonalizing of the matrix $R$.
In Fig. 1 this distribution is given for the value $\omega =0.92$ for which the non-thermal case  level spacing distribution
is chaotic \ci{sal97}. As is seen from this figure for $\theta =0$ it is the same as the results of non-thermal calculations \ci{sal97}.
By increasing the temperature it becomes closer to a Gaussian distribution that means strengthening of chaos in the thermal case.
In Fig. 2 the level spacing distribution for $\omega =0.01$ is plotted. For this value of $\omega$ the systems is
regular at $\theta =0$.  However the increase of temperature  leads to a chaotization of the system and $P(S)$ becomes closer to the
Gaussian distribution.
Figs. 3 and 4 present level spacing distributions (for the same values of parameters as in Figs 1 and 2) for the spectrum calculated
by diagonalizing of the matrix $Z$. By comparing Figs. 1, 2, 3  and 4,  it is clear that at zero temperature level spacing distributions
are the same for
both methods. By increasing the temperature,  the difference between spacing distributions for different methods becomes considerable.
It is clear from Figs. 1 and 2, that the level spacing distribution for  $\omega =0.92$,  $\theta=0.01$ is closer to the Gaussian distribution
compared to ones for   $\omega =0.01$,  $\theta=0.01$. The same behavior can be seen in histograms  for
$\omega =0.92$,  $\theta=0.2$ and $\omega =0.01$,  $\theta=0.2$.
The reason for such  behavior can be understood from Fig.5 where temperature $T$ is plotted as a function of $\theta$ for various
values of  $\omega$.
It is clear that  higher values of $\omega$ correspond to higher temperatures, while for smaller values of  $\omega$
temperature is also small.

However,  for $\omega =0.92$ the difference between the results for $H$ and $\hat H$ is small
while  for $\omega =0.01$ there is considerable difference in the results even for
$\theta=0.01$ (see Fig. 2 and 4).
This indicates that the results from $H$, physical observables and $\hat H$,
generators of symmetry, are quite similar at high temperature
 while at low temperature the results are quite different.
We know that at zero temperature $H$ and $\hat H$ perform different functions. But the results indicate that at high temperature their
chaotic behavior is similar.

Thus in both cases  increasing  the temperature leads to a
smooth transition from a Poissonian to a Gaussian form in the level spacing distribution.
Furthermore, at higher  temperatures both $H$ and $\hat H$ lead to quite similar results.

\section{Conclusion}
Summarizing we have treated quantum chaos in gauge fields at finite temperature
 using a toy model, $SU(2)$ Yang-Mills-Higgs system. To account  for the finite -temperature effects we used
 the thermofield dynamics technique. The need for simultaneous exploration of the   level fluctuations
  and the finite-temperature effects
 is dictated by recent advances in relativistic  heavy ion collision experiments, that allow the creation of hot and dense
 quark-gluon and hadronic matters \ci{QM}.

  The lattice QCD calculations of hadronic matter and quark-gluon matter indicate that both systems exhibit strong chaotic dynamics
\ci{Hal,Berg}. The present calculations in a toy model appear to support such conclusions. Furthermore the study of
the Yang-Mills-Higgs system at finite temperature establishes clearly that an increase of temperature of the system strengthens
level fluctuations in the spectra. Then a transition from the Poissonian to a  Gaussian  level-spacing distribution does occur.
It is to be anticipated that a study of the quark-gluon  system in the relativistic  heavy ion collisions at RHIC and later on at LHC
would show the phenomenon of quantum chaos in a quantum field theoretic system. It is to be emphasized that a
proper study in $(3+1)$ dimensions  with full details of the Yang-Mills non-abelian field along with the Higgs scalar field
is needed to make necessary conclusions for a quark-gluon plasma.
However the present study provides an indication of the possible outcome in more realistic studies.

\section*{Acknowledgments}
The work of DUM is supported  by NATO Science Fellowship of Natural
Science and Engineering Research Council of Canada (NSERCC).
The work of FCK is supported by NSERCC.
The work of URS is supported by a Grant of the  Uzbek Academy of Sciences  (contract No 38-02).
The work of AES is supported by  CNPQ (Brazil).

\end{document}